\documentclass[sigconf,nonacm]{acmart}

\usepackage{amsmath}
\usepackage{booktabs}
\usepackage{cleveref}
\usepackage{microtype}

\setcopyright{none}
\renewcommand\footnotetextcopyrightpermission[1]{}

\newcommand{\SME}{\textsc{SME}}
\newcommand{\SVE}{\textsc{SVE}}
\newcommand{\AoSoV}{\textsc{AoSoV}}
\newcommand{\SoA}{\textsc{SoA}}
\newcommand{\Lvec}{L-vector}
\newcommand{\Svec}{S-vector}
\newcommand{\Evec}{E-vector}
\newcommand{\Lineshine}{Lineshine}
\newcommand{\LXtwo}{LX2}

\begin{document}

\title{Beyond Fast Contractions: Attenuation and Recovery of Matrix-Engine Speedups in High-Order Finite Elements}

\author{Yinuo Wang}
\email{wyn22@mails.tsinghua.edu.cn}
\affiliation{%
  \institution{Tsinghua University}
  \city{Beijing}
  \country{China}}

\author{Lin Gan}
\email{lingan@tsinghua.edu.cn}
\affiliation{%
  \institution{Tsinghua University}
  \city{Beijing}
  \country{China}}
\affiliation{%
  \institution{Hetao Institute of Mathematics and Interdisciplinary Sciences}
  \city{Shenzhen}
  \country{China}}

\author{Tianqi Mao}
\email{mtq24@mails.tsinghua.edu.cn}
\affiliation{%
  \institution{Tsinghua University}
  \city{Beijing}
  \country{China}}

\author{Zeyu Song}
\email{zy-song22@mails.tsinghua.edu.cn}
\affiliation{%
  \institution{Tsinghua University}
  \city{Beijing}
  \country{China}}

\author{Wubing Wan}
\email{wanwubing@himis-sz.cn}
\affiliation{%
  \institution{Hetao Institute of Mathematics and Interdisciplinary Sciences}
  \city{Shenzhen}
  \country{China}}

\author{Jiayu Fu}
\email{fujy24@mails.tsinghua.edu.cn}
\affiliation{%
  \institution{Tsinghua University}
  \city{Beijing}
  \country{China}}

\author{Zekun Yin}
\email{Zekun.yin@sdu.edu.cn}
\affiliation{%
  \institution{Shandong University}
  \city{Jinan}
  \country{China}}

\author{Yuyang Jin}
\email{jinyuyang@tsinghua.edu.cn}
\affiliation{%
  \institution{Tsinghua University}
  \city{Beijing}
  \country{China}}

\author{Xiaohui Duan}
\email{sunrise.duan@sdu.edu.cn}
\affiliation{%
  \institution{Shandong University}
  \city{Jinan}
  \country{China}}

\author{Wei Xue}
\email{xuewei@tsinghua.edu.cn}
\affiliation{%
  \institution{Tsinghua University}
  \city{Beijing}
  \country{China}}

\author{Guangwen Yang}
\email{ygw@tsinghua.edu.cn}
\affiliation{%
  \institution{Tsinghua University}
  \city{Beijing}
  \country{China}}

\begin{abstract}
Modern processors increasingly provide matrix engines whose peak arithmetic
throughput greatly exceeds conventional SIMD, but scientific applications
rarely realize this advantage end to end.  We examine this gap in SPECFEM3D's
dominant stiffness operator on the Arm \LXtwo{} CPUs that power the flagship
\Lineshine{} supercomputer.  Against a matched,
high-performance \SVE{} baseline on the same cores, \SME{}'s $4\times$
single-precision peak advantage falls to $2.2\times$ for isolated tensor
contractions and $1.1\times$ for the complete operator.  Our factorized
diagnostic attributes the loss to pointwise computation, indirect field
movement and synchronization, and irregular coefficient delivery.  Explicit
SIMD mitigates
pointwise work, raising the full-operator speedup to $1.3\times$.  Field-layout
changes mitigate indirect movement and synchronization, while vector-blocked
coefficient streaming reduces irregular-access costs; together
they raise speedup to $1.6\times$ at high order.
A contraction-free control bounds further contraction-only gains at
$1.11$--$1.32\times$.  Realizing matrix-engine performance therefore requires
co-designing the entire operator path, not merely replacing its contraction
kernel.
\end{abstract}

\keywords{Arm SME, spectral element method, high-order finite elements,
matrix-free operators, tensor contractions, data layout, performance analysis}

\maketitle

\section{Introduction}
\label{sec:introduction}


On modern manycore and accelerator-based systems, data movement is often
substantially more expensive than arithmetic in both time and energy.
Matrix-free high-order finite and spectral element methods (FEM/SEM) avoid
assembled sparse matrices.  They exploit tensor-product elements, replacing
assembled-matrix accesses with local computation.  This reduces global memory
traffic while exposing high arithmetic intensity and fine-grained parallelism.
These properties improve accuracy per
computational time and make high-order discretizations important for modern
manycore and accelerator-based systems~\cite{kolev2021,ahmad2021}, including
wave propagation, high-order computational fluid dynamics (CFD), and
atmospheric modeling~\cite{patera1984,komatitsch1999,komatitsch2010,
touhami2022,deville2002}.

On the hardware side, matrix units are a major source of peak floating-point
capability, made prominent by GPU Tensor Cores for dense GEMM.  Recent work
remaps sparse products and broader scientific patterns onto
them~\cite{zachariadis2020tsparse,fan2024dtcspmm,shi2025flashsparse,lu2026mmu}.
In high-order stiffness, sum factorization expresses tensor-product
differentiation and weak-form application as batches of small contractions,
making contraction a natural matrix-unit target~\cite{cui2024,tu2026}.

This apparent hardware--algorithm match is incomplete.  A high-order stiffness
operator is not a single contraction kernel: pointwise physics and metrics,
field gathers and scatters, reconciliation of shared degrees of freedom
(DoFs), and irregular coefficient access surround the contractions.  These
stages can attenuate matrix-unit gains even when the contractions themselves
map efficiently, creating a gap between matrix-unit peak, contraction
performance, and complete-operator speedup.

This gap becomes consequential as matrix units move into
general-purpose multicore CPUs.  Arm's Scalable Matrix Extension (\SME{}) is an
important example.  \Lineshine{} pushes this trend to system scale, sustaining
2.198\,EFLOP/s on HPL as the top-ranked June 2026 TOP500
system~\cite{top5002026} with \LXtwo{} CPUs whose single-precision peak is
dominated by \SME{}~\cite{armSME}.

\begin{figure}[!t]
  \centering
  \includegraphics[width=\columnwidth]{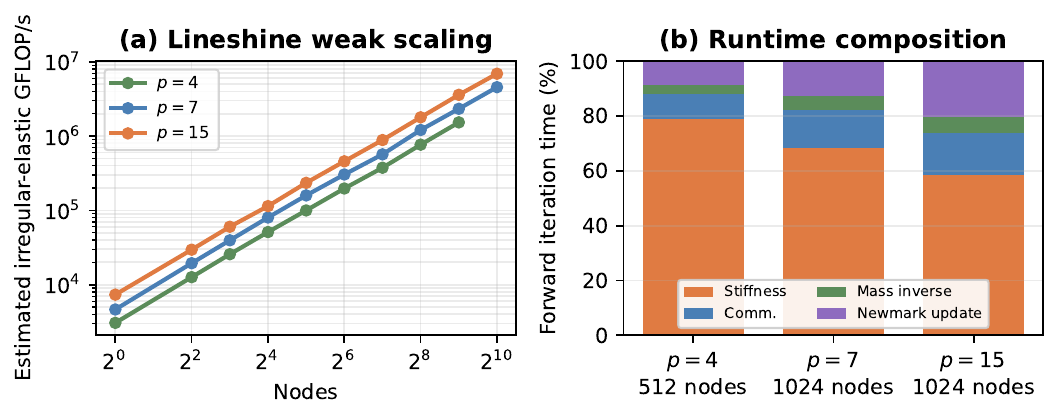}
  \caption{SPECFEM3D weak scaling and solver runtime composition on
  \Lineshine{} for polynomial orders $p\in\{4,7,15\}$.  The application retains
  high parallel efficiency while stiffness remains the dominant measured
  hotspot at scale.}
  \Description{Weak-scaling curves and stacked runtime-composition bars for
  SPECFEM3D runs on Lineshine.}
  \label{fig:lineshine-scaling}
\end{figure}
\begin{figure}[!t]
  \centering
  \includegraphics[width=\columnwidth]{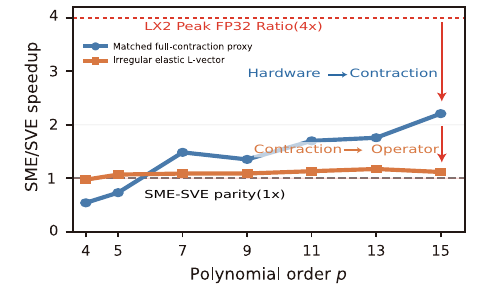}
  \caption{Peak-to-operator attenuation of \SME{} over \SVE{}.  Across
  polynomial order, horizontal lines mark the \LXtwo{} FP32 matrix-FMA peak
  ratio ($4\times$) and parity ($1\times$); the measured curves expose the
  successive loss from matched contractions to the conventional
  compiler-vectorized irregular-elastic \Lvec{} operator.}
  \Description{Line plot of SME-over-SVE speedup across polynomial orders,
  comparing the hardware peak ratio, parity, matched tensor contractions, and
  the conventional irregular-elastic L-vector operator.}
  \label{fig:performance-cascade}
\end{figure}

\Lineshine{} provides a scalable production case: SPECFEM3D retains about
94--98\% weak-scaling efficiency through 512--1024 nodes, with stiffness
consuming 59--79\% of forward-iteration time at the largest scales
(\Cref{fig:lineshine-scaling}).  Yet \SME{}'s $4\times$ FP32 matrix-FMA peak
advantage over Arm's Scalable Vector Extension (\SVE{}) falls to $2.2\times$
for isolated contractions against a matched high-performance \SVE{} proxy.
The full irregular-elastic
\Lvec{}\footnote{We use libCEED terminology~\cite{brown2021libceed}; see
\Cref{sec:background,sec:field-layout} for details.} operator reaches only
$1.1\times$ (\Cref{fig:lvector-nosimd}).  We call this loss from matrix-unit
peak through contractions to full stiffness the
\emph{peak-to-operator attenuation cascade}
(\Cref{fig:performance-cascade}).

This cascade is the starting point of the paper.  Accordingly, we ask:
relative to a matched, high-performance single-instruction, multiple-data
(SIMD) baseline, where is matrix-unit
acceleration exposed or masked across a complete high-order SEM stiffness
operator, and which operator-path and data-layout changes recover it?


To study this cascade, we propose a factorized diagnostic methodology for
high-order FEM stiffness operators that separates operator-level decomposition
from architecture-specific execution backends.  It independently controls the
contraction backend, pointwise physics and geometry implementation, and field
and coefficient layouts while comparing matched null, SIMD, and matrix-unit
contraction backends.  Comparisons among these controlled variants use measured
time, throughput, and effective-bandwidth deltas---rather than
architecture-specific hardware counters---to identify where matrix-unit
acceleration is exposed, masked, or lost.

We instantiate the methodology on \LXtwo{} using \SVE{} and
\SME{} and apply it to SPECFEM3D's acoustic and elastic
formulations~\cite{peter2011specfem}.  Elastic is our primary coupled
vector-field target, while acoustic provides a scalar control for understanding
general FEM stiffness behavior.\footnote{For space, detailed figures focus on
elastic, while the roofline retains both physics.  Complete measurements,
benchmark source code, data, and plotting scripts are available at
\url{https://github.com/FrankTrek/SME_HighOrderFEM_Benchmark}.}  Porting the
methodology to another matrix-engine CPU preserves the same decomposition and
experimental controls while replacing the \SVE{} and \SME{} implementations
with native vector and matrix-unit backends.

The first target is the conventional SPECFEM3D-style \Lvec{} operator.  This
shared-DoF layout is compact and widely used, but surrounds each contraction
with indirect gathers, shared accumulations, and race-avoidance constraints.
The diagnosis first identifies a fixable attenuation layer: pointwise physics
and metric work can hide backend differences unless controlled with explicit
SIMD.  After this fix, attenuation remains in the inherited \Lvec{} structure:
shared DoFs create indirect access and race-avoidance costs, and irregular
geometry adds coefficient and metric traffic that must be separated from
contraction time.

We then use the diagnosis to guide mitigation.  For field movement, \Svec{} and
\Evec{} layouts move shared-DoF costs to different parts of the execution path:
\Svec{} reduces conflicts through subdomain duplication, while \Evec{} exposes
regular element-local kernels at the cost of explicit reconciliation.  The
choice is a tradeoff between indirect access and scatter conflicts inside the
kernel, or explicit synchronization after a more regular local kernel.  For
irregular coefficients, array-of-structures-of-vectors (\AoSoV{}) and
mixed-\AoSoV{} layouts make coefficient access more vector-friendly and reduce
traffic.  In the elastic irregular
case, this recovery path raises the high-order \SME{}/\SVE{} operator speedup
from about $1.1\times$ in the original \Lvec{} diagnostic to about
$1.6\times$ after field and coefficient layout recovery.
The stronger recovered gains at high polynomial order motivate a smooth,
high-fidelity wave-propagation case study of how architecture-dependent
operator throughput can influence equal-accuracy discretization choices.

Taken together, the attenuation diagnosis and layout-mediated recovery reveal
the paper's central insight: matrix-engine acceleration is governed not by the
contraction backend alone, but by where the complete operator pays for
pointwise work, field movement, coefficient delivery, and shared-DoF
reconciliation.

This paper makes four contributions:
\begin{enumerate}
  \item \textbf{Matched \SVE{}/\SME{} contraction backends with an attainable
  performance bound.}  We build strong vector-unit and matrix-unit
  tensor-contraction backends and show that \SME{} approaches the
  order-dependent bound for small SEM contractions.

  \item \textbf{A factorized full-operator diagnosis of matrix-unit
  attenuation.}  Matched null, \SVE{}, and \SME{} backends and controlled
  operator variants expose how pointwise physics, field movement, coefficient
  delivery, and reconciliation attenuate acceleration from contraction to the
  complete operator.

  \item \textbf{Layout-mediated recovery with an operator-level envelope.}
  Explicit SIMD, S/E-vector, and \AoSoV{} variants recover successive
  attenuation layers, reaching about $1.6\times$ high-order elastic-irregular
  \SME{}/\SVE{} speedup; the roofline spans both physics and geometries.

  \item \textbf{A discretization implication for high-order FEM.}  We find
  evidence that the recovered high-order operator throughput favors higher
  polynomial order in the tested equal-accuracy wave-propagation setting.
\end{enumerate}

\section{Background}
\label{sec:background}

\subsection{The LX2 CPU Architecture}

\begin{figure}[t]
  \centering
  \includegraphics[width=\columnwidth]{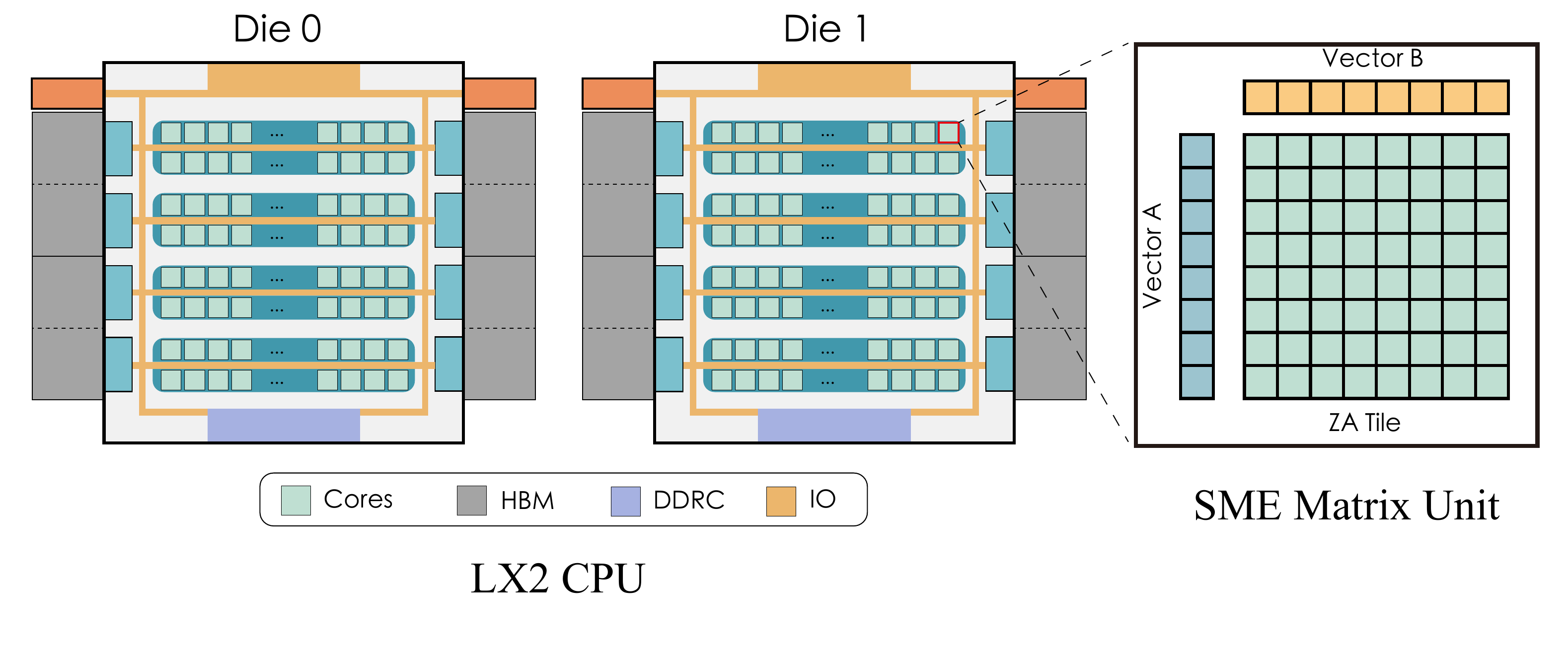}
  \caption{\LXtwo{} organization and per-core \SME{} execution.  The dual-die
  package contains four 38-core NUMA domains per die, connected to HBM stacks,
  DDR controllers, and I/O.  Every CPU core contains an \SME{} matrix unit.
  The inset illustrates the outer product of vectors A and B accumulated into ZA
  using $V_L=8$ for visual clarity (our measured FP32 configuration uses
  $V_L=16$).}
  \Description{Block diagram of the LX2 dual-die package with eight NUMA
  domains, HBM stacks, DDR controllers, and I/O, plus a zoomed per-core SME
  matrix unit.  The inset shows two eight-lane vectors and their outer-product
  accumulation into an eight-by-eight ZA tile.}
  \label{fig:lx2}
\end{figure}

The \LXtwo{} processor used in \Lineshine{} is a dual-die Armv9-A CPU with 304
physical cores running at 1.6\,GHz (\Cref{fig:lx2}).  Each die is divided into
four NUMA domains of 38 cores.  Every core provides conventional \SVE{}/SVE2
vector execution and supports \SME{}, while the memory system combines
per-NUMA high-bandwidth memory (HBM) with socket-level DDR.  Each NUMA domain
provides 4\,GB of HBM, which reaches about 400\,GB/s in our STREAM
measurement~\cite{mccalpin1995stream}, while DDR provides larger capacity at
about 120\,GB/s.

\SME{} evaluates matrix-style computation through an outer-product execution
model.  Two vectors are read from \SVE{} registers, their outer product is
accumulated into the architectural ZA tile, and data moves between ZA and
\SVE{} registers through horizontal or vertical slice instructions.  ZA
consists of 64 rows of 64 bytes, forming a two-dimensional architectural tile,
so practical SEM contractions occupy only part of the tile at low polynomial
order.  For single-precision FMA, \SME{} provides approximately $4\times$ the
theoretical peak throughput of conventional \SVE{} multiply-accumulate
instructions on the same core.  This peak ratio motivates the comparison, but
the rest of the paper studies how much survives once the contraction is
embedded in a full operator.

\begin{figure}[t]
  \centering
  \includegraphics[width=\columnwidth]{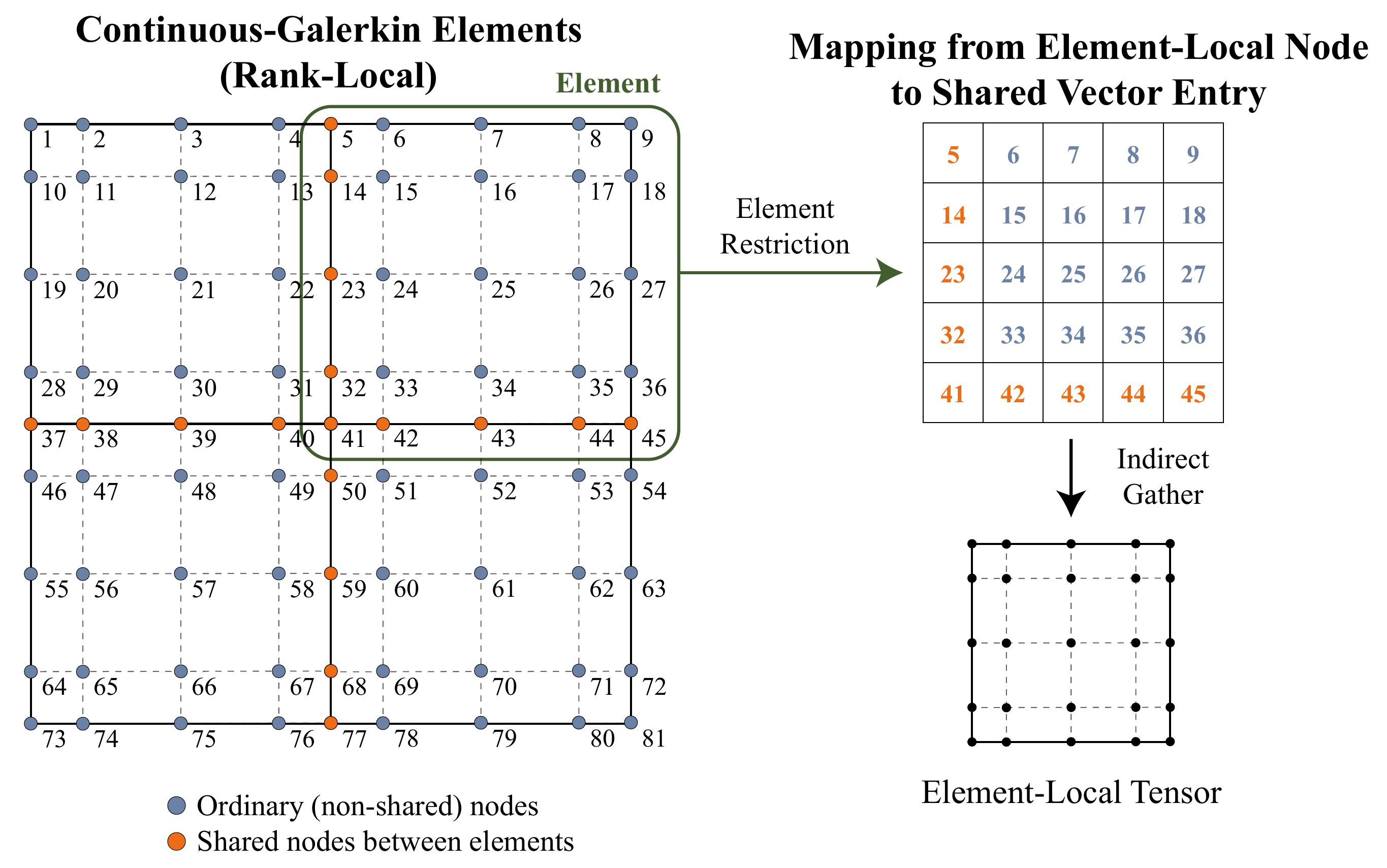}
  \caption{Conventional \Lvec{} storage and element restriction.  A rank-local
  continuous-Galerkin vector assigns one entry to each distinct DoF, with
  interface entries shared by neighboring elements.  An indirect restriction
  map gathers these entries into a dense element-local tensor for operator
  evaluation.}
  \Description{Rank-local continuous-Galerkin L-vector with ordinary and
  element-shared nodes, and an indirect restriction map gathering shared
  entries into an element-local tensor.}
  \label{fig:lvector-layout}
\end{figure}

\subsection{High-Order FEM and Tensor Contraction}

The stiffness operators arise from weak forms of acoustic and elastic wave
propagation.  For scalar acoustic propagation, we use
\begin{equation}
  a_{\mathrm{ac}}(u,v)
  = \int_{\Omega} \rho^{-1} \nabla v \cdot \nabla u \,\mathrm{d}x ,
  \label{eq:acoustic}
\end{equation}
where $u$ and $v$ are scalar trial and test fields and $\rho^{-1}$ is inverse
density.  For elastic propagation, we use
\begin{equation}
  a_{\mathrm{el}}(\mathbf{u},\mathbf{v})
  = \int_{\Omega}
    \nabla \mathbf{v} : \mathbb{C} : \nabla \mathbf{u}
    \,\mathrm{d}x ,
  \label{eq:elastic}
\end{equation}
where $\mathbf{u}$ and $\mathbf{v}$ are displacement trial and test fields and
$\mathbb{C}$ is the elastic stiffness tensor.  In the isotropic case used here,
$\mathbb{C}$ is parameterized by the Lam\'e parameters $\lambda$ and $\mu$.

Each physical hexahedral element $\Omega_e$ is represented by a map
$\mathbf{x}_e(\xi,\eta,\zeta)=(x,y,z)$ from the reference cube
$\widehat{\Omega}=[-1,1]^3$; its tensor-product
Gauss--Lobatto--Legendre (GLL) nodes lie at
$(\xi_i,\eta_j,\zeta_k)$.  SEM evaluates gradients through this tensor-product
structure.  At polynomial order $p$, let $n=p+1$ denote the number of GLL
points in each reference-coordinate direction.  The element-local field is a small
three-dimensional tensor $u_{ijk}$, with $i,j,k=0,\ldots,n-1$.  Let $D$ denote
the one-dimensional differentiation matrix evaluated at the GLL points.  The
three reference-space derivatives differ only in the tensor dimension to
which $D$ is applied; using $\xi$ as representative,
\begin{equation}
  (\partial_\xi u)_{ijk}
    = \sum_{a=0}^{n-1} D_{ia} u_{ajk}.
  \label{eq:derivatives}
\end{equation}
The $\eta$ and $\zeta$ forms apply $D$ to the second and third indices,
respectively.  Each directional derivative is a tensor contraction,
equivalently a batch of small matrix multiplications over element-local
slices~\cite{canuto2007}.  The test-function gradients enter through a
transpose-gradient operation with the same tensor-product structure using
$D^T$.

Between the gradient and transpose-gradient contractions, the element kernel
performs pointwise physics and geometry work.  At each quadrature point, a
$3\times3$ contravariant matrix maps reference-space derivatives to physical
space.  The acoustic factor $\rho^{-1}$ or the elastic tensor $\mathbb{C}$ is
applied.  The resulting physical flux is mapped back to reference space
before the transpose-gradient forms the element-local residual.

The \Lvec{} maps each element-local GLL node to a shared vector entry
(\Cref{fig:lvector-layout}).  Its compact storage requires indirect
gather/scatter, while concurrent updates require atomics or conflict-free
element coloring~\cite{Jones1993}.

Restriction, directional contractions, pointwise physics and geometry,
transpose contractions, and accumulation define the attenuation path studied
below.  \Cref{sec:contraction} calibrates the matrix-unit opportunity in the
contractions.  \Cref{sec:lvector} separates pointwise physics, field movement,
and coefficient traffic in the conventional \Lvec{} operator.
\Cref{sec:layout-recovery} changes field and coefficient representations to
recover the exposed losses.

\begin{figure*}[!t]
  \centering
  \includegraphics[width=\textwidth]{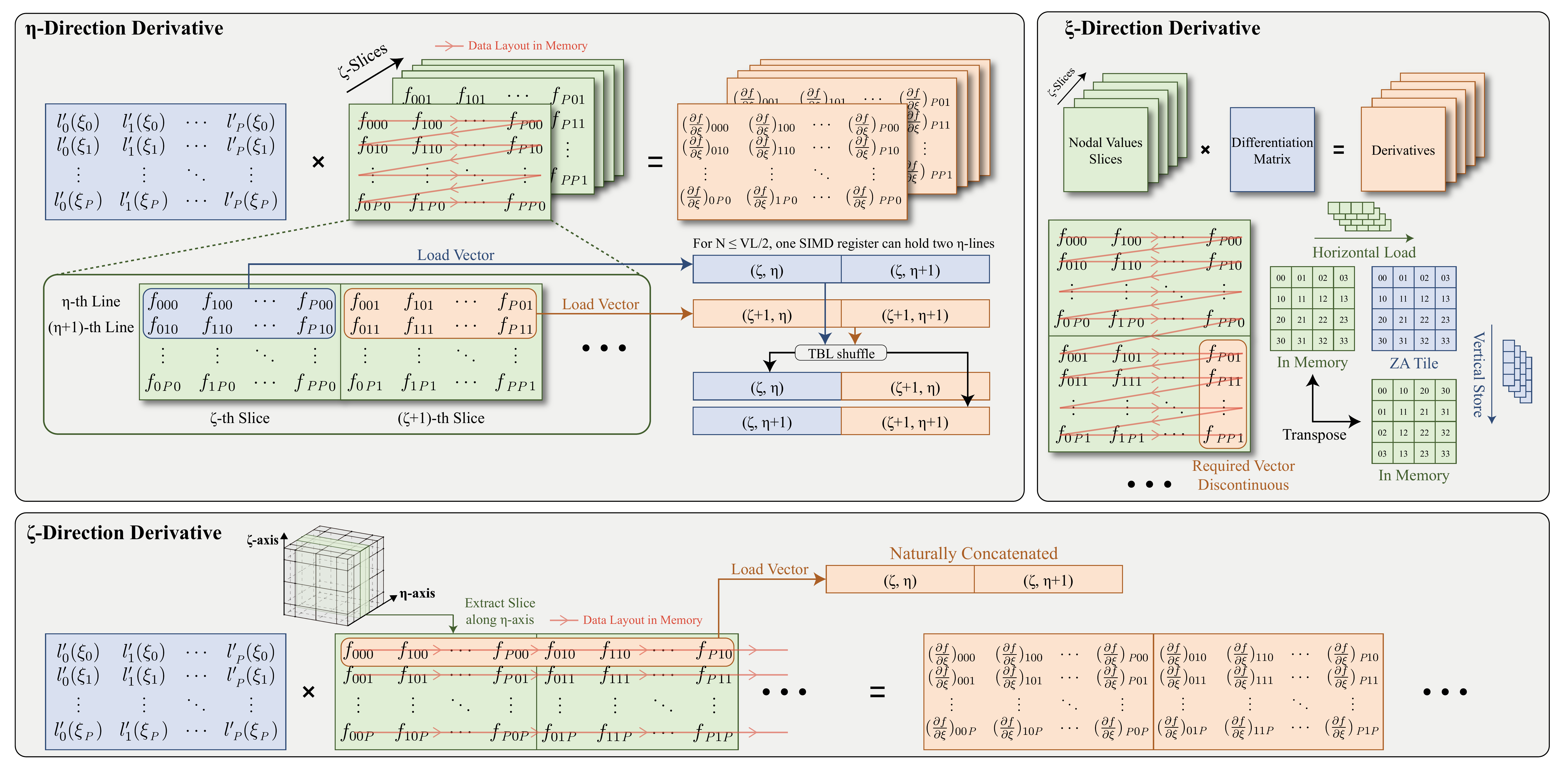}
  \caption{Direction-specific data movement in the \SME{} contraction backend.
  The $\eta$ path packs short lines, the $\xi$ path reorganizes discontinuous
  memory operands through ZA-tile transposition, and the $\zeta$ path exploits
  naturally contiguous slices.}
  \Description{Diagrams of memory layout, vector loading, shuffling,
  transposition, and ZA-tile organization for eta-, xi-, and zeta-direction
  derivatives.}
  \label{fig:sme-backend}
\end{figure*}

\begin{figure*}[!t]
  \centering
  \includegraphics[width=\textwidth]{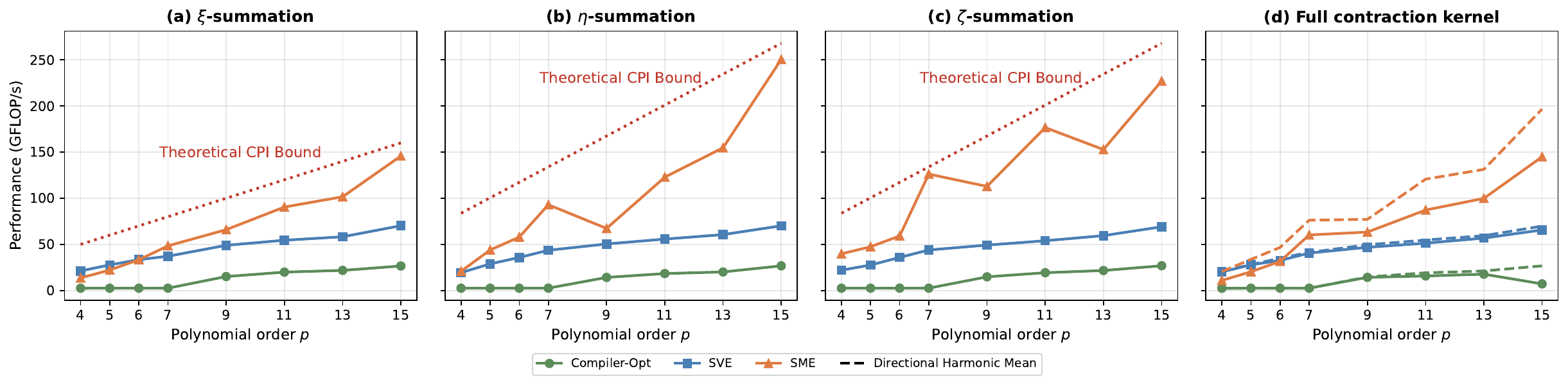}
  \caption{Single-core directional tensor contractions and full-contraction
  proxy.  The \SME{} curves should be compared with the order- and
  direction-dependent attainable bounds rather than nominal matrix-FMA peak.}
  \Description{Performance curves over polynomial order for compiler, SVE, and
  SME directional contractions and a complete contraction proxy.}
  \label{fig:tensor-contraction}
\end{figure*}

\section{Experimental Setup, Validation, and Data Policy}
\label{sec:setup}

All measurements use the vendor-provided \LXtwo{} platform.  Each run is
confined to one HBM NUMA domain and uses 36 CPU cores, with HBM holding the hot
stiffness working set.  OpenMP threads are pinned to physical cores with
\texttt{OMP\_PLACES=cores} and \texttt{OMP\_PROC\_BIND=close}; each timed
invocation follows 50 warm-up iterations with 200 measured iterations.  We
report FP32 unless otherwise stated.  \SVE{} and \SME{} binaries are built from
matched kernel contracts, and every optimized path is validated through
canonical operator outputs against the native flat-\Lvec{} reference
($10^{-4}$ absolute and relative tolerances).  Mixed-\AoSoV{} storage is
accepted only when relative $L_2$ and $L_\infty$ errors are each at most
$10^{-3}$.  Builds use \texttt{clang++} with \texttt{-mcpu=hip11},
\texttt{-O3}, \texttt{-ffast-math}, and a fixed 512-bit \SVE{} vector length.

The explicit-SIMD design-space sweep provides 16 process-level samples for
each of 840 configurations, yielding 13,440 timings; the compiler-vectorized
control uses five.  A scripted stability check identifies 246 samples
(1.83\%) as intermittent high-side slowdowns.  These samples are sparse in
most configurations and are excluded before taking the median.  For the few
problematic configurations, the slowdowns affect a substantial fraction of
repetitions or exhibit collection-cohort dependence, making filtering alone
unreliable.  We therefore remeasure these configurations using unchanged
binaries and matched null, \SVE{}, and \SME{} commands executed sequentially
on one dedicated HBM NUMA domain.  We report the median of five complete
process-level repetitions without further filtering, retaining a point only
when all outputs validate.

The artifact repository contains original and rerun measurements, command
lines, binary hashes, sweep manifests, filtering code, validation tolerances,
NUMA placement, OpenMP settings, compiler flags, thresholds, exclusions, and
backend build macros.  The manuscript reports the selected summary points
needed for the performance narrative.

\section{Contraction-Layer Attenuation: Bound and Matched Backends}
\label{sec:contraction}

The first attenuation layer appears before the full stiffness operator is
considered. Both backends evaluate the directional gradient and transpose-gradient
contractions represented by \Cref{eq:derivatives}: \SVE{} uses unrolled,
register-tiled SIMD FMA kernels, while \SME{} maps the work to outer-product
updates on ZA tiles.  Despite \SME{}'s higher peak, SEM contractions are small.
Each directional derivative acts on length-$n$ lines inside an
$n\times n\times n$ cube, so each line exposes only $n$ useful contraction
opportunities.  Before $n$ approaches the \SME{} vector/tile width $V_L$, the
attainable utilization of matrix-FMA peak scales as $n/V_L$.  This is a shape
bound imposed by the operator, not a code-generation artifact.

Load/store overhead further lowers the attainable rate.  For the $\eta$- and
$\zeta$-direction contractions, data can be aggregated into \SME{}-friendly
operands, but accumulated results still have to be stored.  We model the
first-order bound as
\begin{equation}
  \frac{\mathit{CPI}_{\mathrm{OPA}}}
       {\mathit{CPI}_{\mathrm{OPA}}+\mathit{CPI}_{\mathrm{LSU}}}
  \cdot \frac{n}{V_L}\,\mathit{FLOPS}_{\mathrm{SME}}.
  \label{eq:sme-bound}
\end{equation}
Here $\mathit{CPI}$ denotes cycles per instruction; $\mathrm{OPA}$ and
$\mathrm{LSU}$ denote outer-product accumulation and the load/store unit,
respectively.
The $\xi$ direction has an additional layout penalty because element-local
memory order does not match the operand layout required by outer products.
The kernel performs in-register transpose and reorganization to avoid strided
loads.  This adds roughly one extra load and store stream, lowering the bound
to
\begin{equation}
  \frac{\mathit{CPI}_{\mathrm{OPA}}}
       {\mathit{CPI}_{\mathrm{OPA}}+3\mathit{CPI}_{\mathrm{LSU}}}
  \cdot \frac{n}{V_L}\,\mathit{FLOPS}_{\mathrm{SME}}.
  \label{eq:sme-xi-bound}
\end{equation}
For the 512-bit FP32 configuration, $V_L=16$ lanes and the rounded
single-core matrix-FMA peak is
$\mathit{FLOPS}_{\mathrm{SME}}=400$\,GFLOP/s.  The modeled issue costs
$\mathit{CPI}_{\mathrm{OPA}}=2$ and $\mathit{CPI}_{\mathrm{LSU}}=1$ yield
attainable-rate factors of $2/3$ for $\eta,\zeta$ and $2/5$ for $\xi$.

Our backend is designed to approach these direction-specific bounds
(\Cref{fig:sme-backend}).  It aggregates work across the batched slice
dimension and pipelines independent slices across ZA tiles.  For the $\zeta$
direction, element layout exposes contiguous $\xi\eta$ planes, so the kernel
aggregates them into a larger $n\times n^2$ multiplication.  The $\xi$
direction uses in-register transpose/reorganization to recover contiguous
vector access.  The $\eta$ direction is closest to a direct batched small
matrix multiplication over $\zeta$ slices; for $p=4$--$7$, it additionally
uses vector concatenation and shuffle operations to pack multiple short
element-local lines into one outer-product operand.  The matched \SVE{} path
ensures that later measurements compare hardware datapaths rather than a
strong \SME{} implementation against a weak SIMD baseline.

\Cref{fig:tensor-contraction} reports single-core microbenchmarks.  At high
order, the \SVE{} backend reaches about 65\,GFLOP/s, roughly 65\% of the
single-core \SVE{} FMA peak; given the small matrix sizes and pipeline and
load/store overheads, it is a strong vector-engine baseline.  The directional
results show that the CPI-style bound is the right point of comparison.
Performance grows as $n$ occupies more of ZA, while the $\eta$ and $\zeta$
curves retain visible store overhead and $\xi$ remains lower because of
transpose/reorganization.  The $\eta$ direction drops after $p=7$, where the
low-order concatenation path no longer applies, but resumes the same
tile-utilization trend at higher orders.

Across directional kernels, \SME{} reaches up to 250\,GFLOP/s and outperforms
\SVE{} by up to $3.6\times$ at high order.  At $p=15$, the speedups are
$2.1\times$, $3.6\times$, and $3.3\times$ for $\xi$, $\eta$, and $\zeta$,
respectively.  Compared with the compiler-optimized scalar backend, the
directional \SME{} kernels are one order of magnitude faster at high order.

\begin{figure*}[t]
  \centering
  \includegraphics[width=\textwidth]{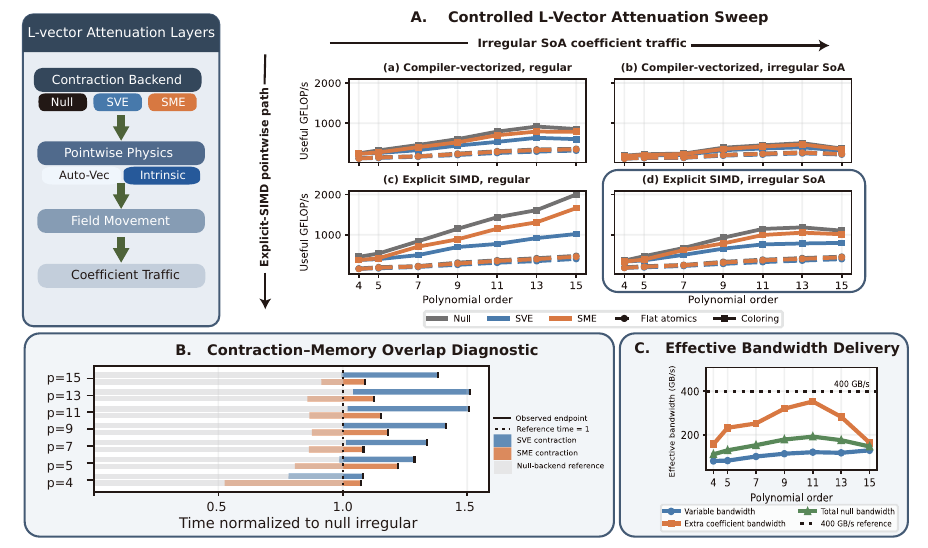}
  \caption{Factorized diagnosis of \Lvec{} attenuation for elastic stiffness.
  The sidebar traces contraction backend, pointwise SIMD, field movement, and
  irregular \SoA{} coefficient traffic.  (A) Controlled backend sweeps
  separate SIMD and geometry effects.  (B) The partial-overlap analysis
  compares observed, null-backend, and estimated contraction costs.  (C)
  Effective-bandwidth measurements expose the residual shared-DoF and
  coefficient-streaming limitations.}
  \Description{Composite figure with a four-layer attenuation diagram,
  controlled elastic L-vector throughput sweeps, a contraction-memory overlap
  diagnostic, and an effective-bandwidth diagnostic across polynomial order.}
  \label{fig:lvector-diagnostic-composite}
  \label{fig:lvector-nosimd}
  \label{fig:lvector-simd}
  \label{fig:lvector-overlap}
  \label{fig:lvector-bandwidth}
\end{figure*}

The full contraction proxy is the relevant calibration for stiffness.  Because
all three directions execute in sequence, its effective rate is governed by
the slower direction through a harmonic-mean effect.  At $p=15$, the
full-proxy \SME{}/\SVE{} speedup is about $2.2\times$, close to the $\xi$
speedup and well below $\eta$ and $\zeta$.  The compiler-optimized backend is
not a reliable comparison target: it changes behavior sharply across orders
and exhibits a cliff at $p=15$.  The gap between the directional harmonic mean
and full proxy also grows at high order, consistent with the working set
exceeding the 32\,KB L1 cache.  Thus a $4\times$ hardware peak advantage
becomes about $2.2\times$ at the contraction layer even before full-operator
costs are introduced.

\section{L-Vector Attenuation Diagnosis}
\label{sec:lvector}

\subsection{Diagnostic Design and Metrics}

The benchmark preserves the matrix-free operator while omitting file, mesh-I/O,
and orchestration infrastructure so implementations can vary independently.
Its axes are contraction backend (null, \SVE{}, \SME{}), regular or irregular
geometry, scalar acoustic or vector elastic physics, and compiler-vectorized
or explicit-SIMD pointwise work.  Together they separate contraction,
variable movement, physics, and coefficient traffic.

Let $T_{\mathrm{null}}$ denote null-backend stiffness time and $T_b$ the
measured time for $b\in\{\mathrm{SVE},\mathrm{SME}\}$.  The modeled hot-path
traffic is split into variable-field traffic $Q_{\mathrm{var}}$ and
coefficient/metric traffic $Q_{\mathrm{coeff}}$.  We use three diagnostic
quantities:
\begin{enumerate}
  \item \textbf{Variable-field bandwidth.}  For regular geometry, the null
  backend gives the control
  $B_{\mathrm{var}}=Q_{\mathrm{var}}/T_{\mathrm{null}}^{\mathrm{reg}}$.

  \item \textbf{Incremental coefficient bandwidth.}  For irregular geometry,
  the irregular-minus-regular delta estimates pointwise-stream cost as
  $B_{\mathrm{coeff}}^{\mathrm{extra}}
  =Q_{\mathrm{coeff}}^{\mathrm{irreg}}/
  (T_{\mathrm{null}}^{\mathrm{irreg}}-T_{\mathrm{null}}^{\mathrm{reg}})$.
  This is an effective incremental quantity, not raw copy bandwidth: the
  timing subtraction can include overlap with variable movement and
  arithmetic.

  \item \textbf{Total null-backend bandwidth.}  As a conservative aggregate,
  we report $B_{\mathrm{null}}^{\mathrm{total}}
  =(Q_{\mathrm{var}}+Q_{\mathrm{coeff}})/T_{\mathrm{null}}$.
\end{enumerate}

To reason about why \SVE{} or \SME{} does not simply add contraction time to
the null time, we use a partial-overlap model.  Let
$\widehat{T}_{\mathrm{contr}}^b$ be the contraction time estimated from
\Cref{sec:contraction}.  We write
\begin{equation}
  T_b \approx
  \max(T_{\mathrm{mem}},\widehat{T}_{\mathrm{contr}}^b)
  +\alpha_b
  \min(T_{\mathrm{mem}},\widehat{T}_{\mathrm{contr}}^b),
  \label{eq:overlap}
\end{equation}
where $T_{\mathrm{mem}}$ is null-backend memory/control time and $\alpha_b$ is
a fitted non-overlap parameter.  Values near zero indicate good overlap;
values near one indicate near serialization.  The model is diagnostic rather
than cycle accurate.

\begin{table}[t]
  \centering
  \footnotesize
  \setlength{\tabcolsep}{2pt}
  \caption{Elastic irregular-\SoA{} \Lvec{} workload.  All cases use $h=0.5$;
  mesh dimensions vary to keep the HBM-resident problem scale comparable.
  Footprint includes field, numbering, coefficient, plan, reference-element,
  and scratch storage.  Work is analytical GFLOP per application, and
  contraction share is a modeled FLOP fraction rather than measured time.}
  \Description{Mesh dimensions, element and degree-of-freedom counts, HBM
  footprint, analytical work, and modeled contraction share across polynomial
  order for elastic irregular L-vector stiffness.}
  \label{tab:lvector-workload}
  \begin{tabular*}{\columnwidth}{@{\extracolsep{\fill}}r r r r r r r@{}}
    \toprule
    $p$ & Mesh & $N_e$ (k) & MDoF & GiB & GF/app. & Contr. (\%) \\
    \midrule
     4 & $63{\times}63{\times}56$ & 222.3 & 43.21 & 2.02 &  8.28 & 60.4 \\
     5 & $54{\times}54{\times}44$ & 128.3 & 48.69 & 2.08 &  9.26 & 64.7 \\
     7 & $42{\times}42{\times}36$ &  63.5 & 66.05 & 2.43 & 13.20 & 70.9 \\
     9 & $36{\times}33{\times}28$ &  33.3 & 73.51 & 2.55 & 15.90 & 75.3 \\
    11 & $27{\times}27{\times}24$ &  17.5 & 70.60 & 2.33 & 16.63 & 78.5 \\
    13 & $24{\times}24{\times}20$ &  11.5 & 76.71 & 2.47 & 19.66 & 81.0 \\
    15 & $24{\times}21{\times}16$ &   8.1 & 82.48 & 2.59 & 22.92 & 83.0 \\
    \bottomrule
  \end{tabular*}
\end{table}

\Cref{tab:lvector-workload} uses 2.02--2.59\,GiB elastic irregular-\SoA{}
working sets, chosen to fit within one NUMA domain's HBM while providing enough
independent element work to occupy all 36 cores and measure steady-state,
resource-saturated performance.  From $p=4$ to $15$, contraction length grows
from $n=5$ to $16$ and modeled FLOP share from 60.4\% to 83.0\%, marking the
transition from short-line, larger-pointwise-share cases to full-width,
contraction-heavier cases.  These shares are not runtime fractions.

\subsection{Attenuation in the L-Vector Path}

We apply the \Lvec{} attenuation diagnosis to both physics.  For elastic,
\Cref{fig:lvector-diagnostic-composite} shows the complete factorized
diagnosis for our primary coupled-vector target.  Acoustic follows the same
attenuation sequence, while its simpler pointwise map provides the scalar
control; we quantify this contrast below.

\textbf{Compiler-vectorized.}  In \Cref{fig:lvector-nosimd}, elastic irregular
reaches only 187--401\,GFLOP/s with \SVE{} and 182--470\,GFLOP/s with \SME{},
giving 0.97--$1.17\times$ \SME{}/\SVE{}---far below the $2.2\times$
contraction proxy.  Coloring is $1.09$--$3.59\times$ faster than flat atomics
and is therefore the \Lvec{} baseline.

\textbf{Explicit SIMD.}  For acoustic, explicit-SIMD/no-SIMD performance spans
0.98--$1.18\times$ in regular geometry and 0.97--$1.03\times$ in irregular
geometry, peaking at $1.18\times$.  Its simpler pointwise map therefore shows
substantially less mitigation than elastic.  For elastic irregular, explicit
SIMD improves coloring by
$1.77$--$2.52\times$ for \SVE{} and $1.81$--$2.86\times$ for \SME{}
(\Cref{fig:lvector-simd}).  The high-order \SME{}/\SVE{} ratio rises from about
$1.1\times$ to $1.27\times$, but remains well below the contraction proxy.
This distinction between absolute throughput and relative backend speedup is
important: explicit SIMD removes a pointwise-physics bottleneck shared by both
backends, but does not restore contraction to the dominant share of operator
time.

\textbf{Residual attenuation.}  For elastic irregular, estimated contraction
time is only 0.18--$0.55\times$ null time, while observed \SME{} time is
1.07--$1.22\times$ null time (\Cref{fig:lvector-overlap}).  Irregular \SoA{}
also raises elastic null time by $1.4$--$2.0\times$; variable-field
bandwidth is 80--129\,GB/s and incremental coefficient bandwidth is
156--353\,GB/s, both below the 400\,GB/s STREAM reference
(\Cref{fig:lvector-bandwidth}).  These values do not indicate saturated bulk
bandwidth.  Rather, coloring, indirect shared-DoF access, and fragmented
coefficient streams prevent the operator from using STREAM-like delivery.
Together, the overlap and bandwidth diagnostics identify structural field and
coefficient movement---not contraction speed alone---as the remaining loss.

\begin{figure*}[t]
  \centering
  \includegraphics[width=\textwidth]{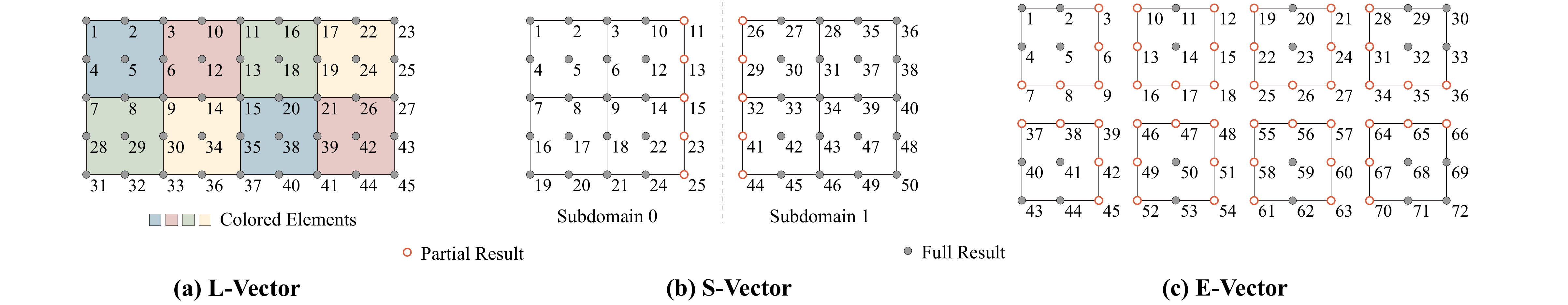}
  \caption{Field storage and accumulation.  \Lvec{} uses shared DoFs with
  colored execution; \Svec{} and \Evec{} duplicate interface DoFs by subdomain
  and element, respectively.  Storage indices remain contiguous across each
  rank-local vector.  Partial interface results require reconciliation,
  whereas locally complete results do not.}
  \Description{Three layouts with continuously numbered nodes: shared
  L-vector storage with colored elements, subdomain-private S-vector storage,
  and element-local E-vector storage.  Markers distinguish locally complete
  results from partial interface results requiring reconciliation.}
  \label{fig:field-layouts}
\end{figure*}

\section{Data-Layout Recovery Across the Full Operator}
\label{sec:layout-recovery}

\begin{figure*}[t]
  \centering
  \includegraphics[width=\textwidth]{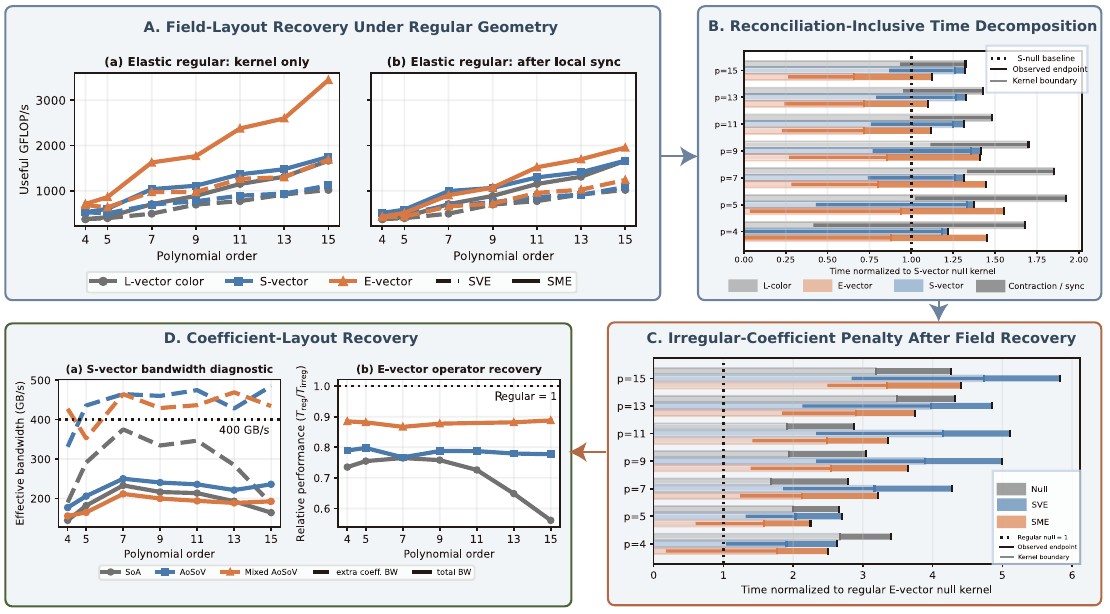}
  \caption{Field- and coefficient-layout recovery for elastic stiffness.
  (A) Regular-geometry \Lvec{}, \Svec{}, and \Evec{} results isolate field
  locality.  (B) Kernel and reconciliation-time decomposition shows where
  shared-DoF work moves.  (C) Reintroducing irregular \SoA{} coefficients
  isolates the remaining traffic penalty.  (D) \Svec{} bandwidth diagnostics
  and reconciliation-inclusive \Evec{} results show recovery from \AoSoV{} and
  mixed-\AoSoV{} storage.}
  \Description{Four-panel composite showing regular field-layout recovery,
  kernel and reconciliation time, the irregular-coefficient penalty after
  field recovery, and coefficient-layout bandwidth and operator recovery.}
  \label{fig:layout-mitigation}
  \label{fig:field-recovery-composite}
  \label{fig:regular-layout-progression}
  \label{fig:field-coeff-time}
  \label{fig:coefficient-recovery-composite}
  \label{fig:svec-coeff-bandwidth}
  \label{fig:evec-coeff-ratio}
\end{figure*}

After explicit SIMD exposes movement costs, we recover full-operator
performance in two stages: first field locality, then coefficient delivery.
\Cref{fig:layout-mitigation} summarizes this complete recovery sequence.
Unless marked kernel-only, reported results include reconciliation.

\begin{figure}[t]
  \centering
  \includegraphics[width=\columnwidth]{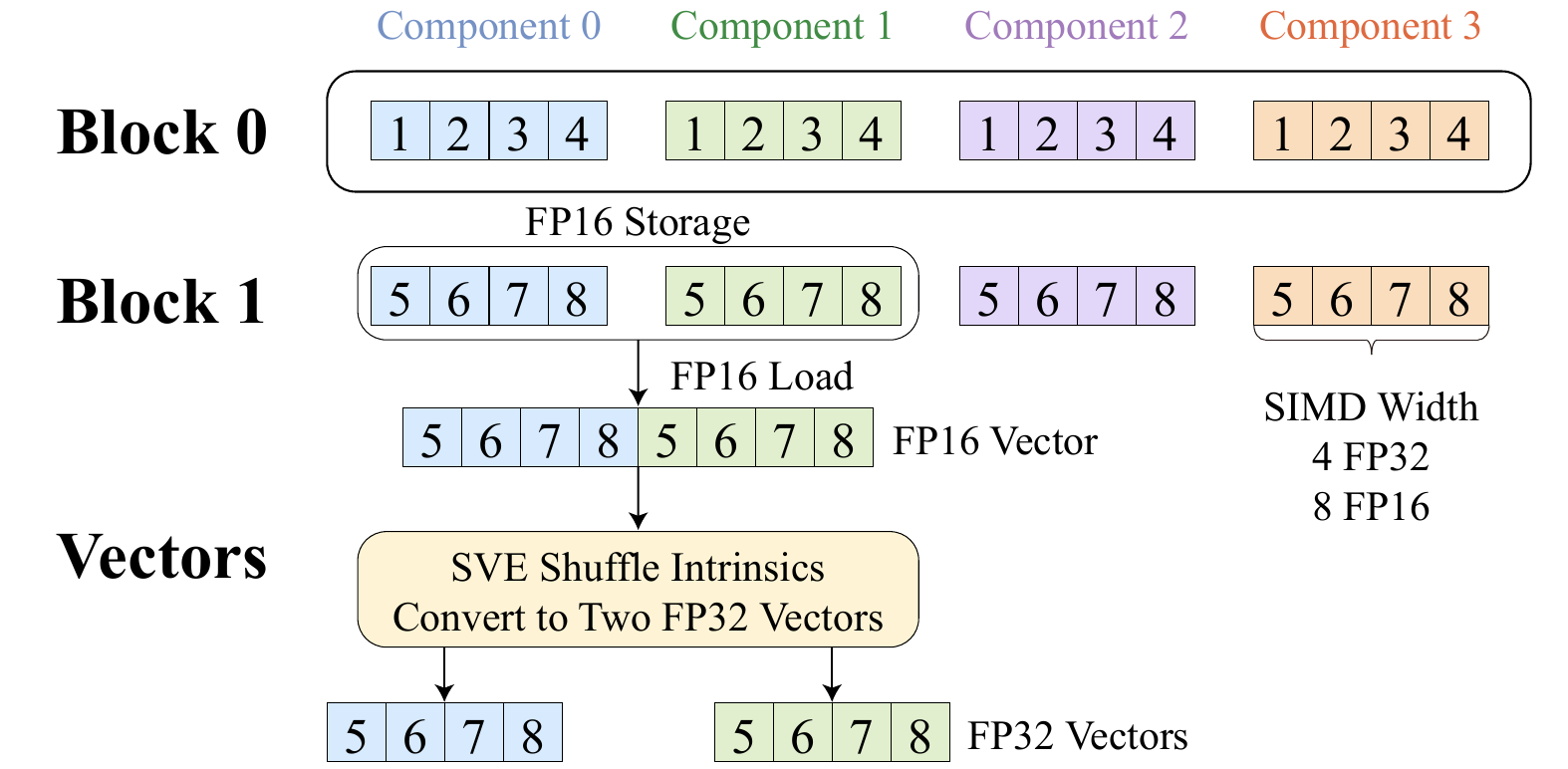}
  \caption{\AoSoV{} coefficient semantics and mixed-precision conversion.
  Each block stores one SIMD width of points contiguously by component.  Mixed
  \AoSoV{} packs selected components in FP16, then shuffles and converts each
  load into two FP32 vectors for pointwise computation.}
  \Description{Coefficient components organized into SIMD-width AoSoV blocks,
  followed by the mixed-AoSoV path from one packed FP16 load through SVE
  shuffling to two FP32 vectors.}
  \label{fig:aosov-concept}
\end{figure}

\subsection{Field-Layout Recovery: L-, S-, and E-Vectors}
\label{sec:field-layout}

\Cref{fig:field-layouts} defines the \Lvec{}/\Svec{}/\Evec{} semantics; panels
A--B of \Cref{fig:layout-mitigation} present elastic results.  Acoustic follows
the same locality sequence as a lower-field-traffic scalar control; all
complete-operator results include reconciliation.

\Svec{} uses first-touch, thread-private subdomains with duplicated boundaries
and local reconciliation.  \Evec{} stores every DoF per element for contiguous,
conflict-free kernels, then reconciles copies to canonical DoFs.  Both trade
in-kernel indirection and conflicts for duplication and explicit synchronization.

\textbf{(A) Field locality.}
Starting from explicit-SIMD \Lvec{} coloring, elastic \SME{} at $p=15$
improves from 1663\,GFLOP/s in \Lvec{} to 1748\,GFLOP/s in \Svec{} and
3447\,GFLOP/s in kernel-only \Evec{}.  Field reorganization raises absolute
throughput for both backends rather than uniquely favoring \SME{}.

\textbf{(B) Reconciliation cost.}
Including reconciliation reduces \Evec{} to 1954\,GFLOP/s: its
$1.97\times$ kernel advantage over \Svec{} narrows to $1.12\times$ for the
complete operator.  Element-local storage therefore moves, rather than
eliminates, much of the shared-DoF work.

\subsection{Coefficient-Layout Recovery: AoSoV}
\label{sec:coeff-layout}

After field-layout recovery, irregular geometry leaves coefficient delivery as
the next attenuation layer.  We evaluate this stage for both physics: panels
C--D of \Cref{fig:layout-mitigation} present elastic results, while acoustic
provides the scalar control.

\SoA{} stores each coefficient component as a separate element-major stream.
\AoSoV{} instead follows the vector-blocked AoSoA
principle~\cite{duan2017saligner}, grouping one SIMD width of GLL points with
block-contiguous components (\Cref{fig:aosov-concept}).  Mixed \AoSoV{} packs
selected coefficients in FP16, then converts them for the same FP32 formulas
and contraction backend.

\textbf{(C) Exposed coefficient penalty.}
With field access already element-local, switching from regular data to
irregular \SoA{} coefficients produces a $1.78\times$
reconciliation-inclusive slowdown at $p=15$.  This isolates metric and
material streams as the next attenuation layer.  Because the contiguous
\Evec{} null path perturbs streaming, panel D uses the more stable \Svec{} null
backend for bandwidth diagnosis.

\textbf{(D) Coefficient-traffic recovery.}
\AoSoV{} raises effective coefficient delivery above the 400\,GB/s STREAM
reference for $p=5$--$15$; this overlap-aware metric is not literal sustained
bandwidth.  At $p=15$, reconciliation-inclusive \Evec{} performance relative
to regular geometry improves from 0.56 with \SoA{} to 0.78 with \AoSoV{} and
0.89 with mixed \AoSoV{}.  The latter uses validated FP16 coefficient storage,
not a general accuracy guarantee for arbitrary distorted meshes.  For
acoustic, the \AoSoV{}/\SoA{} effective-bandwidth ratio spans
0.91--$1.35\times$ across orders and reaches $1.35\times$ at $p=15$,
confirming that high-order vector-blocked coefficient delivery also benefits
scalar stiffness.  Thus field locality exposes the coefficient bottleneck,
and vector-blocked coefficient storage recovers most of the remaining
irregular-geometry penalty.

\subsection{Best-Configuration Envelope and Roofline Placement}
\label{sec:envelope}

\begin{figure}[t]
  \centering
  \includegraphics[width=\columnwidth]{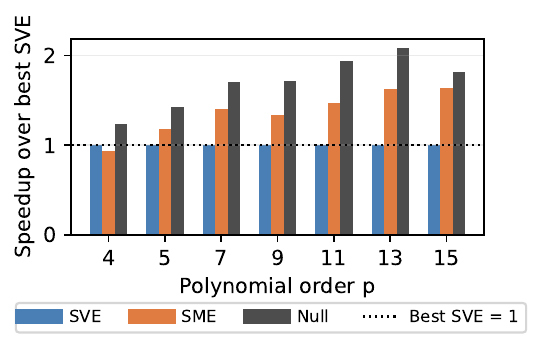}
  \caption{Best validated elastic-irregular speedup envelope.  Each backend
  selects its fastest pointwise implementation, field/coefficient layout, and
  execution plan; bars are normalized to the best \SVE{} configuration at
  each order.}
  \Description{Bar chart of SVE-normalized speedup for SVE, SME, and null on
  the elastic irregular operator across polynomial order.}
  \label{fig:backend-envelope}
\end{figure}

\begin{figure}[t]
  \centering
  \includegraphics[width=\columnwidth]{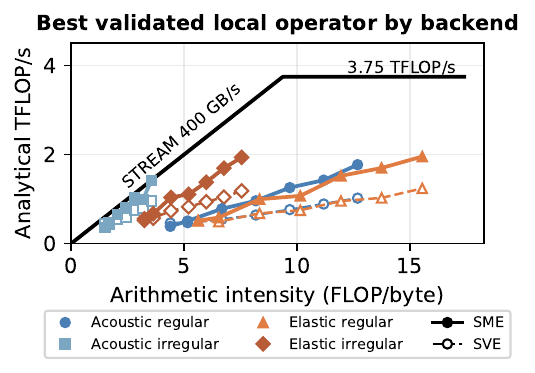}
  \caption{Characteristic roofline after layout recovery, with intensity fixed
  to analytical work and the original traffic model.  The horizontal compute
  roof is the 3.75\,TFLOP/s \SVE{} peak.  Filled solid and open dashed curves
  denote the best validated configurations for \SME{} and \SVE{},
  respectively.}
  \Description{Roofline plot of the best validated SVE and SME configurations
  for acoustic and elastic regular and irregular stiffness.}
  \label{fig:best-roofline}
\end{figure}

The elastic-irregular envelope spans \Lvec{}, \Svec{}, and \Evec{} fields;
compiler-vectorized and explicit-SIMD pointwise physics; \SoA{}, \AoSoV{}, and
mixed-\AoSoV{} coefficients; and available execution and reconciliation
plans.  Invalid or tolerance-failing results are excluded, and \Svec{}/\Evec{}
times include reconciliation.  Each backend independently selects its fastest
remaining configuration at every order, so \Cref{fig:backend-envelope} is a
best-to-best rather than fixed-layout comparison.

\begin{figure*}[t]
  \centering
  \includegraphics[width=\textwidth]{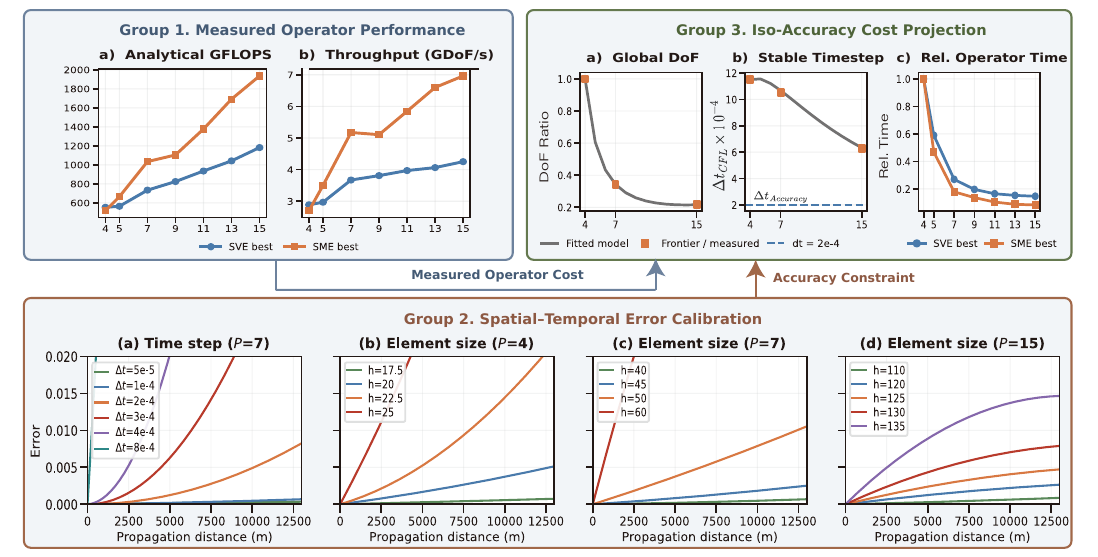}
  \caption{From measured operator throughput to iso-accuracy stiffness cost.
  Group 1 reports reconciliation-inclusive elastic-irregular GFLOP/s and
  GDoF/s.  Group 2 uses station-wise waveform errors and fitted trends to
  determine the largest element size satisfying the normalized squared
  waveform $L_2$-error threshold $10^{-2}$ at each polynomial order.  Group 3
  combines global DoF count, stable time step, and measured operator time to
  expose the projected stiffness-cost tradeoff.}
  \Description{Three-group composite containing two measured operator
  throughput panels, four waveform-error calibration panels, and three panels
  for global DoFs, stable time step, and relative operator time.}
  \label{fig:discretization-implications-composite}
  \label{fig:elastic-throughput}
  \label{fig:fit-error-trend}
  \label{fig:accuracy-cost}
\end{figure*}

The envelope exposes the corresponding order-dependent transition.  At
$p=4$, best \SME{} reaches only $0.94\times$ best \SVE{}, while even the null
backend reaches only $1.23\times$.  The short $n=5$ contraction occupies
$5/16$ of the FP32 matrix width and accounts for 60.4\% of modeled FLOPs, so
pointwise physics, coefficient loads and stores, field gathers and scatters,
and reconciliation remain prominent.  As order increases, fuller matrix
utilization and a rising contraction share generally raise \SME{}/\SVE{} to
$1.64\times$ at $p=15$, increasing the original irregular \Lvec{} result from
about $1.1\times$ to about $1.6\times$.

Yet null/\SME{} remains only 1.11--$1.32\times$ across the sweep and is
$1.11\times$ at $p=15$.  The null backend removes contraction instructions
while preserving the surrounding operator path, so this ratio is both a
diagnostic of non-contraction cost and an empirical counterfactual ceiling
from making contractions free.  Thus contraction FLOP share describes the
increasing opportunity for matrix acceleration, whereas the null envelope
measures the remaining elapsed-time benefit.  This is not a strict hardware
bound because removing contraction can perturb caching and overlap.

\Cref{fig:best-roofline} places configurations selected under the same
best-validated rule on the characteristic
roofline~\cite{williams2009roofline}, keeping intensity tied to analytical
work and the original \Lvec{} traffic model so layout recovery appears as
vertical progress for both \SVE{} and \SME{}.  Across acoustic and elastic,
regular and irregular operators, recovered \SME{} curves sit above the
corresponding \SVE{} curves at comparable intensity.  Acoustic irregular
\SME{} reaches 80.5\% of the \SVE{} roof with FP32 \AoSoV{} and 99.8\% with
mixed \AoSoV{} at $p=15$, so its near-roofline result additionally depends on
reduced coefficient precision.  Elastic irregular \SME{} reaches 56.2\% and
64.1\% of the \SVE{} roof with FP32 and mixed \AoSoV{}, respectively.  Its
qualitative recovery therefore does not depend on mixed precision, while
remaining farther from the roof than the acoustic cases.

\section{Discretization Implications}
\label{sec:discretization}

The recovered \SME{} configurations attain their highest analytical FLOP rates at
high order, but tensor-product stiffness performs $O(n^4)$ work for $O(n^3)$
unknowns, where $n=p+1$.  We therefore connect arithmetic rate to DoF
throughput and an iso-accuracy cost projection in
\Cref{fig:discretization-implications-composite}.

\textbf{(1) Operator throughput.}
Reconciliation-inclusive GDoF/s rises toward high order for both backends
(\Cref{fig:elastic-throughput}).  Improved hardware utilization therefore
offsets the increasing work per DoF, showing that the high-order GFLOP/s gain
is not merely an arithmetic-count artifact.

\textbf{(2) Iso-accuracy frontier.}
We consider dispersion-dominated error for a homogeneous scalar 20\,Hz Ricker
wave problem with Newmark time stepping.  For receiver $s$, we compare complete
observed and reference waveforms over $[0,T_{\mathrm{rec}}]$ using
\begin{equation}
  \varepsilon_s =
  \frac{\int_0^{T_{\mathrm{rec}}}
    [f_{\mathrm{obs},s}(t)-f_{\mathrm{ref},s}(t)]^2\,\mathrm{d}t}
  {\int_0^{T_{\mathrm{rec}}} f_{\mathrm{ref},s}(t)^2\,\mathrm{d}t}.
  \label{eq:waveform-error}
\end{equation}
We require the normalized squared waveform $L_2$ error
$\varepsilon_s\leq10^{-2}$.  Group 2 first varies
$\Delta t=5\times10^{-5}$--$8\times10^{-4}$ at $p=7$ and $h=40$; we select
$\Delta t_{\mathrm{acc}}=2\times10^{-4}$ where temporal error is comparable
to spatial error.  To isolate the spatial frontier, element-size sweeps at
$p=4,7,15$ use $\Delta t=10^{-4}$, with $h\in\{17.5,20,22.5,25\}$,
$\{40,45,50,60\}$, and $\{110,120,125,130,135\}$, respectively.  Error tables
sample propagation distance from 0.5 to 12.49\,km at 10\,m intervals.  For
each sweep, station-wise error versus distance is fit by least squares to
$c_1x+c_2x^2$.  At each $p$, the frontier uses the largest tested $h$ whose
fitted error at the receiver corresponding to approximately 10\,s of
propagation satisfies the criterion (\Cref{fig:fit-error-trend}).  This
selects $(p,h)=(4,20),(7,50),(15,125)$.  We use this homogeneous scalar-wave
calibration as a relative points-per-wavelength proxy for the elastic
projection, assuming its order-to-order spatial-resolution trend transfers to
smooth elastic propagation.  Relative to $p=4$, the latter points reduce the
three-dimensional DoF count to 0.343 and 0.216, respectively, and also reduce
pressure on the 4\,GB-per-NUMA HBM capacity.

\textbf{(3) Projected stiffness cost.}
We combine the frontier with measured elastic-irregular throughput as follows,
where $\Delta t_{\mathrm{CFL}}$ denotes the
Courant--Friedrichs--Lewy (CFL) stability limit:
\begin{equation}
  T_{\mathrm{stiff}}^{\mathrm{proj}}
  \propto
  \left\lceil
    \frac{T_{\mathrm{phys}}}
    {\min(\Delta t_{\mathrm{acc}},\Delta t_{\mathrm{CFL}})}
  \right\rceil
  \frac{N_{\mathrm{DoF}}}{R_{\mathrm{DoF}}}.
  \label{eq:stiffness-cost}
\end{equation}
Here $R_{\mathrm{DoF}}$ is reconciliation-inclusive throughput and the step
count is set by the smaller of the accuracy- and CFL-admissible time steps.
Because minimum GLL spacing scales as $O(h/n^2)$, increasing $h$ competes with
the high-order CFL penalty.  The measured CFL limit decreases from
$1.15\times10^{-3}$ at $p=4$ to $1.05\times10^{-3}$ at $p=7$ and
$6.3\times10^{-4}$ at $p=15$.

The accuracy-selected step, $2\times10^{-4}$, is below all three limits, so
projected stiffness speedup equals per-step speedup.  From $p=4$ to $p=7$ it is
$3.71\times$ for \SVE{} and $5.58\times$ for \SME{}.  From $p=7$ to $p=15$ it
is $1.84\times$ and $2.14\times$.  Under a CFL-limited
sensitivity these become $3.38\times$ and $5.10\times$, then $1.10\times$ and
$1.28\times$.  Thus $p=7$ is favorable in either regime, whereas the added
benefit of $p=15$ depends on time-step selection
(\Cref{fig:accuracy-cost}).  These are stiffness-contribution projections, not
end-to-end speedups.

This frontier is deliberately a smooth, homogeneous wave-propagation case
study rather than a universal prescription for high order.  Realistic
simulations add geometric error from curved elements, heterogeneous material
interfaces, and mesh-quality constraints; shocks, under-resolved features, and
nonlinear stability can further require filtering or regularization.  These
effects are established concerns in high-order
methods~\cite{vos2010htop,ainsworth2004dispersion,ainsworth2009spectral,
wang2013highorder,klockner2011shock} and may shift the hardware-favored
frontier.

\section{Related Work}
\label{sec:related}

\paragraph{Matrix Engines for Scientific Operators.}
Matrix engines now support a widening set of scientific operators beyond dense
GEMM, including sparse--sparse multiplication~\cite{zachariadis2020tsparse},
sparse--dense and sampled matrix products~\cite{fan2024dtcspmm,shi2025flashsparse},
stencil computations~\cite{liu2022tcstencil,chen2024convstencil,
gu2026tensorstencil,zhang2024lorastencil,wang2026smestencil,
huang2025hstencil}, and broader
parallel patterns~\cite{lu2026mmu}.

High-order FEM is a natural target because sum factorization exposes tensor
contractions.  LIBXSMM accelerates small CPU products~\cite{heinecke2016};
GPU work optimizes tensor-product kernels~\cite{swirydowicz2019} and maps them
to tensor cores~\cite{cui2024}.  Concurrent work by Tu et al.\ programs FP64
tensor cores for fused full-operator MFEM kernels at scale, including a tsunami
digital twin~\cite{tu2026,henneking2025}.

\paragraph{Limits of Prior Evidence and Our Position.}
Prior evidence consists largely of microkernel studies or
implementation-to-implementation comparisons between matrix-unit and separate
CUDA-core, SIMD, or library paths.  Reported gains may therefore combine
matrix instructions with changes in data reuse, work mapping, and kernel
fusion, making it difficult to attribute speedup specifically to the matrix
unit.  Coverage, data movement, and numerical format further govern application
benefit~\cite{domke2021,lu2026mmu}.  We instead compare matched \SVE{}, \SME{},
and null backends on the same \LXtwo{} cores, then factor pointwise SIMD, field
representation, reconciliation, and coefficient layout across the complete
operator.

Our contribution is an operator-level methodology rather than a new framework.
Mainstream systems including libCEED~\cite{brown2021libceed},
MFEM~\cite{anderson2021mfem,andrej2024}, deal.II~\cite{arndt2021},
Nek5000/NekRS~\cite{fischer2022nekrs}, and
SPECFEM3D~\cite{peter2011specfem} expose relevant matrix-free abstractions.
Their optimized paths, however, do not always make the contraction backend,
pointwise implementation, field representation, coefficient layout, and
reconciliation strategy independently selectable.  Our methodology identifies
independent control of these interfaces as a key framework-level requirement
for isolating matrix-engine benefit and recovering acceleration across the
complete operator.

\section{Conclusion}
\label{sec:conclusion}

Useful matrix-engine speedup is a property of the whole operator path.  On
\LXtwo{}, the $4\times$ \SME{} peak advantage becomes about $2.2\times$ for
matched contractions and about $1.1\times$ in the conventional irregular
\Lvec{} operator.  Coordinated pointwise, field-, and coefficient-layout
recovery raises the high-order elastic-irregular speedup to about
$1.6\times$.

Four operator-level lessons emerge.  First, explicit SIMD for pointwise
physics is necessary to expose backend differences that compiler-dependent
vectorization can mask.  Second, the null backend is only
$1.11$--$1.32\times$ faster than recovered \SME{}, providing an empirical
ceiling on further contraction-only optimization.  Third, continuous-Galerkin
shared-DoF cost is relocated rather than eliminated: \Lvec{}, \Svec{}, and
\Evec{} trade in-kernel indirection and conflicts against storage duplication
and reconciliation.  Finally, vector-blocked \AoSoV{} aligns coefficient
delivery with SIMD consumption; its FP32 recovery shows that the qualitative
benefit does not depend on mixed precision.

The measured ratios are \LXtwo{}-specific, but the diagnostic is not: whenever
a matrix unit accelerates only the contraction subset of an operator,
pointwise work, field movement, coefficient delivery, and reconciliation
jointly determine the realizable speedup.  Exposing matrix units therefore
requires full-operator representation contracts rather than an isolated fast
contraction.  The smooth wave-propagation case further suggests that recovered
throughput can favor higher polynomial order, while extension to realistic
geometries, heterogeneous media, and nonlinear problems remains future work.

\bibliographystyle{ACM-Reference-Format}
\bibliography{references}

\end{document}